\documentclass[conference, 9pt]{IEEEtran}
\IEEEoverridecommandlockouts
\usepackage{amsmath,amssymb,amsfonts}
\usepackage{algorithm}
\usepackage{algpseudocode}
\usepackage{graphicx}
\usepackage{textcomp}
\usepackage{xcolor}
\usepackage{subfigure}
\usepackage[numbers]{natbib}
\usepackage{flushend}
\usepackage{multirow}
\usepackage{array}
\usepackage{lipsum}
\usepackage{capt-of}
\usepackage{cuted} 
\usepackage{booktabs}

\def\BibTeX{{\rm B\kern-.05em{\sc i\kern-.025em b}\kern-.08em
    T\kern-.1667em\lower.7ex\hbox{E}\kern-.125emX}}
\begin{document}

\title{Deep Sylvester Posterior Inference for Adaptive Compressed Sensing in Ultrasound Imaging 
\thanks{This work was supported by the European Research Council (ERC) under the ERC starting grant nr. 101077368 (US-ACT). 

We thank SURF (www.surf.nl) for their support in using the Dutch National Supercomputer Snellius.}
}

% \author{
% \IEEEauthorblockN{
% Simon Penninga\IEEEauthorrefmark{1},
% Hans van Gorp, 
% Ruud van Sloun}
% \IEEEauthorblockA{
% \textit{Eindhoven University of Technology, Department of Electrical Engineering, Eindhoven, Netherlands} \\
% correspondence: s.w.penninga@tue.nl\IEEEauthorrefmark{1}
% }}
\author{\IEEEauthorblockN{Simon W. Penninga}
\IEEEauthorblockA{\textit{Department of Electrical Engineering} \\
\textit{Eindhoven University of Technology}\\
Eindhoven, The Netherlands \\
s.w.penninga@tue.nl}
\and
\IEEEauthorblockN{Hans van Gorp}
\IEEEauthorblockA{\textit{Department of Electrical Engineering} \\
\textit{Eindhoven University of Technology}\\
Eindhoven, The Netherlands \\
h.v.gorp@tue.nl}
\and
\IEEEauthorblockN{Ruud J.G. van Sloun}
\IEEEauthorblockA{\textit{Department of Electrical Engineering} \\
\textit{Eindhoven University of Technology}\\
Eindhoven, The Netherlands \\
r.j.g.v.sloun@tue.nl}
}

\maketitle
\noindent
\begin{abstract}
Ultrasound images are commonly formed by sequential acquisition of beam-steered scan-lines.
Minimizing the number of required scan-lines can significantly enhance frame rate, field of view, energy efficiency, and data transfer speeds. Existing approaches typically use static subsampling schemes in combination with sparsity-based or, more recently, deep-learning-based recovery. In this work, we introduce an adaptive subsampling method that maximizes intrinsic information gain \textit{in-situ}, employing a Sylvester Normalizing Flow encoder to infer an approximate Bayesian posterior under partial observation in real-time. Using the Bayesian posterior and a deep generative model for future observations, we determine the subsampling scheme that maximizes the mutual information between the subsampled observations, and the next frame of the video. We evaluate our approach using the EchoNet cardiac ultrasound video dataset and demonstrate that our active sampling method outperforms competitive baselines, including uniform and variable-density random sampling, as well as equidistantly spaced scan-lines, improving mean absolute reconstruction error by 15\%. Moreover, posterior inference and the sampling scheme generation are performed in just 0.015 seconds (66Hz), making it fast enough for real-time 2D ultrasound imaging applications.
\end{abstract}

\vspace{2mm}


\section{Introduction}
Ultrasound systems perform sequences of pulse-echo experiments, called transmit events, to form an image. Due to the physical speed of sound, the optimization of these transmit events constitutes a trade-off between frame rate, depth of view, and image quality, making acquisition time a major limiting resource. By reducing the amount of transmit events required to form an image, the effective budget one can spend on this trade-off improves greatly. In addition, subsampling can reduce data transfer and battery drain, enabling cheaper and more portable ultrasound systems. 

Efficient subsampling and signal recovery can be achieved with Compressed Sensing \cite{Donoho2006CompressedSensing}, in which sparsity in some signal domain is used for effective reconstruction from compressed measurements, such as an undersampled Fourier spectrum and or observations from sparse arrays. Contemporary recovery methods go beyond signal sparsity and use deep learning to exploit the signal structure learned from the training data. In particular, deep generative models explicitly learn signal priors that can subsequently be used for inference. Such approaches have also been used in the context of ultrasound imaging \cite{Stevens2024DehazingModels, Asgariandehkordi2024DenoisingModels}. 
Deep learning also enables the optimization of the subsampling schemes themselves \cite{Wang2023DeepTesting}. For instance, Deep Probabilistic Subsampling (DPS) \cite{Huijben2020LearningImaging} uses an end-to-end deep learning training method that finds the optimal subsampling strategy for downstream recovery tasks. 
Additional examples include subsampling of RF data through deep learning \cite{Yoon2019EfficientLearning} and randomized channel subsampling for increased ultrasound speeds \cite{Yu2023RandomizedImaging}.
However, the aforementioned methods have in common that the learned subsampling masks are fixed, and their optimization does therefore not benefit from any information gained across the sequential sampling process at inference time. Conversely, adaptive sensing methods exploit previously acquired data to optimize future sampling schemes across a sequence of observations to improve performance \cite{VanGorp2021ActiveSubsamplingb}. 

\begin{figure*}[ht]
    \centering
    \includegraphics[width=\linewidth]{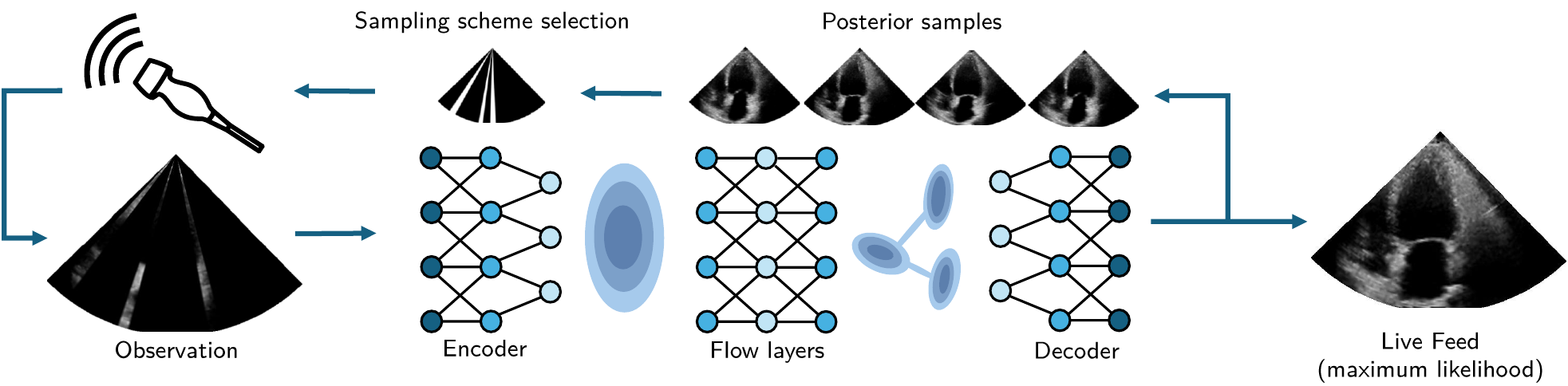}
    \caption{Schematic overview of the active sampling loop of a single video frame. Partial observations of the full frame are used to estimate the latent posterior distribution of the next frame of the video. Samples from this posterior distribution are used to estimate mutual information between the state and the observation, which in turn determines the next subsampling mask and results in new observations.}
    \label{fig:intro}
\end{figure*}

In this paper, we propose an active subsampling method for ultrasound imaging that: (1) exploits a deep generative latent variable model and combines it with a deep Bayesian posterior encoder that performs fast inference of the parameters of its approximate latent posterior from partial observations; (2) designs adaptive subsampling schemes that maximize information gain \textit{on the fly} across a sequence of ultrasound image frames in a video. Specifically, we optimize the evidence lower bound and train a deep neural network to estimate the parameters of the intricate latent posterior state distribution under partial observations, which we parameterize using a Sylvester normalizing flow \cite{Berg2018SylvesterInference}. Based on samples from this posterior, we subsequently design a new sampling scheme that optimizes an estimate of the expected information gain, by maximizing the marginal entropy for future observations. Both steps of the approach are illustrated in Fig.~\ref{fig:intro}.

%This implementation allows us to address the optimization of the acquisition scheme through the minimization of free energy \cite{Buckley2017TheReview}, linking it to the principles of active inference \cite{Friston2017TheInference}, an explanation for biological intelligence stemming from the field of neuroscience. 
%A similar approach has been suggested in `Deep Adaptive Design` \cite{Foster2021DeepDesign} and the `Cognitive Radar` \cite{Haykin2012CognitiveEngineering}, but to the best of our knowledge, these ideas were left without a functioning implementation at scale.

Most related to our approach, van de Camp \textit{et al.} recently proposed the use of deep generative latent variable models for adaptive subsampling designs
\cite{vandeCamp2023ActiveGain}. While effective, the method relied on Markov Chain Monte-Carlo methods for generating samples from the posterior, rendering inference prohibitively slow for time-sensitive applications such as ultrasound imaging. Moreover, the scene was considered static, and observations of this static scene were taken one at a time. In contrast, we here operate on sequences of ultrasound frames and design full subsampling masks for each next frame sequentially. Using the Sylvester normalizing flow-based posterior encoder (requiring only a single neural function evaluation), we reduce inference time by several orders of magnitude, enabling real-time processing rates, while retaining the ability to fit intricate posteriors.

%Performance will be assessed with a line-scanning angle selection simulation, in which the system controls the subset of directions for which lines are transmitted.

The remainder of this paper is organized as follows; Section \ref{sec:methods} describes the problem setup for ultrasound line-scanning, our approach to fast posterior inference, and the design of sampling schemes based on mutual information. In Section \ref{sec:exp} the method is applied to sequences of ultrasound frames, and compared against non-adaptive baselines. Finally, in Section \ref{sec:disc_and_concl}, we conclude and outline future work. 
\section{Methods}
\label{sec:methods}
\subsection{Problem setup}
Let a partial observation of a video frame at a given time-step $y_t$ be defined as:
\begin{equation}
    y_t = A_t x_t + n_t,
    \label{eq:yaxn}
\end{equation}
where $A_t\in \mathbb{R}^{M\times N}$ is the binary subsampling matrix (with $M<<N$), $x_t\in \mathbb{R}^{N}$ is the fully-sampled vectorized video frame that we will refer to as the \textit{image state}, and $n_t$ is the added noise at the time of observation $t \in [0, 1, ..., T]$. The goal is to (1) perform efficient estimation of the Bayesian posterior for image states $p(x_t|y_t, A_t)$, and (2) design an optimal future sampling matrix $A_{t+1}$ that maximizes expected information gain.

%Unfortunately, the population of real-world image states $x_t \in \chi$ is generally too sparse with respect to its dimensionality $\mathbb{R}^N$ to perform any tractable form of Bayesian inference.
Unfortunately, computing the true Bayesian posterior quickly turns intractable in high dimensions. To overcome this, we use a deep latent variable model that approximates the true distribution of signals $p(x)\approx \int p_\theta(x|z)p(z)dz$ using a simplified and lower-dimensional latent distribution $p(z)$, with $z\in \mathbb{R}^{D_z}$ and $D_z<<N$. Our goal then becomes to infer $p(z_t|y_t)$. When confronted with strongly subsampled image states and highly ambiguous observations, i.e. $M<<N$, this posterior will nevertheless remain intricate and often multi-modal. 

\subsection{Deep Sylvester Posterior Inference}
 To model the complex distribution $p(z_t|y_t)\approx q_\phi(z_t|y_t)$, we use a Sylvester Normalizing Flow (Sylvester-NF). The model architecture is an extension of the Variational Auto Encoder (VAE) \cite{Kingma2013Auto-EncodingBayes} and consists of a convolutional image encoder and decoder. The encoder $q_\phi$ outputs the latent Gaussian distribution parameters $\mu(y_t) \in \mathbb{R}^{D_z}$, $\sigma(y_t) \in \mathbb{R}^{D_z}$ and Normalizing Flow \cite{Rezende2015VariationalFlows} parameters $\lambda \in \mathbb{R}^{N_p}$ for subsequent transformations of the Gaussian distribution, with $N_p$ the parameters of the normalizing flow layers. To train the image encoder, we minimize its variational free energy.
 Given an observation $\hat{y}_{t}$ we can formulate the variational free energy as:
\begin{equation}
    \begin{split}
    &-\mathcal{F}(\theta,\phi; \hat{y}_{t}) = \mathbb{E}_{ q_\phi(z_t^{0:K}|\hat{y}_t)}
    \Big[ 
    \underbrace{\log p_\theta(y_{t}=\hat{y}_t|z_t^K)}_{\text{Likelihood }  \textit{Dec}_\theta} - \\ & 
    \underbrace{\log q_\phi(z^0_t|\hat{y}_t)}_{\text{Likelihood } \textit{Enc}_\phi} - \underbrace{\log p(z_t^K)}_{\text{Likelihood }z_t^K}    + \underbrace{\sum_{k=1}^K \log \Big| \text{det}\left(J[z_t^k,\lambda_k(\hat{y}_t)]\right)\Big|}_{\text{LogDet Jacobian Flow transformations}}\Big],  
    \end{split}
    \label{eq:VFE-Sylvester}
\end{equation}
in which $z^0_t \in \mathbb{R}^{D_z}$ is a sample drawn from $\mathcal{N}(\mu(\hat{y}_t),\sigma(\hat{y}_t))$ and $z^K_t \in \mathbb{R}^{D_z}$ the same sample warped through $k \in [1, ..., K]$ flow layers. 
Here, $J$ denotes the Jacobian matrix and $\lambda_k$ the transform parameters of layer $k$. Note that we leave the dependency on the subsampling mask that generates the observations $\hat{y}_t$ implicit throughout this paper. 

The generative latent variable model is first pre-trained using a dataset of full observations $\hat{x}_t\in \chi$ to capture the signal prior $\int p_\theta (x_t|z_t^K)p(z_t^K)dz_t^K$, after which the weights $\theta$ are frozen and the inference model $q_\phi (z_t^K|y_t)$ can be trained for a dataset of partial observations.

% The analytical posterior at a given timestep $t$ is given as as:
% \begin{equation}
% \begin{split}
% \label{eqn_dbl_x}
% &p(x,y_{t:T},A_{t:T},z_{t:T}|y_{1:t},A_{1:t}) =  \\
%      &\frac{p(x,y_{1:t},y_{t:T},A_{1:t},A_{t:T},z)}{\int\int\int \sum_{A_{t+1}}^Tp(x,y_{1:t},y_{t:T},A_{1:t},A_{t:T},z)dzdy_{t:T}dx}.
% \end{split}
% \end{equation}

% Solving (\ref{eqn_dbl_x}) evaluates a path of actions with respect to every other possible path for all time steps from $t$ to $T$. This creates three challenges that need to be addressed in order to make the problem tractable. 
% Firstly, the action horizon $T$ requires iterative updating of the posterior for candidate observations. 
% For now, we choose a `greedy` policy, set $T$ to 1, and leave the exploration of a longer action horizon to future work.

% Second, the integrals in the denominator have to be approximated by a sample aggregate of the posterior from which the generative model creates the expected observations $\hat{y}_{t:T}$. The number of samples drawn affects both performance and resource usage.
% Lastly, the number of possible sampling schemes is generally `n choose k`, which leads to combinatorial scaling and makes the calculation of the sum of all $A_{t+1}$ intractable. 
% To that end, we restrict the possible sampling schemes to an a priori set of possible sampling schemes $S_A$.

\subsection{Sampling Policy}
Our sampling policy is to maximize the information gain of future observations, which is equivalent to minimizing the expected posterior entropy \cite{Foley2024BayesianEntropy}. 
The action-conditional mutual information between future latent states $z_{t+1}$ and observations $y_{t+1}$ for a greedy (one-step-ahead) policy is given by: 
\begin{equation}
    I(y_{t+1};z_{t+1}|A_{t+1},\hat{y}_t) = H(y_{t+1}|A_{t+1},\hat{y}_t) - H(y_{t+1}|z_{t+1},A_{t+1}).
    \label{eq:infmax}
\end{equation}
We leave the exploration of a longer action horizon to future work.
Since the entropy of expected observations $y_{t+1}$ given $z_{t+1}$ does not depend on $A_{t+1}$ (it depends only on the noise $n_{t+1}$), our policy reduces to the maximization of the marginal entropy as:
\begin{equation}
    A_{t+1}^* = \arg\max_{A_{t+1}}\left[ H(y_{t+1}|A_{t+1},\hat{y}_t) \right].
    \label{eq:actionselection}
\end{equation}
The marginal entropy scales with the log determinant of the covariance matrix $\Sigma_{y_{t+1}|A_{t+1}}$, which we estimate using the generative model $p_\theta(x_t|z_t^K)$ and a sample aggregate of the posterior $q_\phi(z_{t+1}^K|\hat{y}_t)$, assuming an identity transition $z_{t+1}^K|z_{t}^K$:
\begin{equation}
\begin{split}
    &\Sigma_{y_{t+1}|A_{t+1},\hat{y}_t} \\
    &= \mathbb{E}_{q_\phi(y_{t+1}^K|\hat{y}_t,A_{t+1})}\left[ (y_{t+1} - \mu_{y_{t+1}})(y_{t+1} - \mu_{y_{t+1}})^T\right] \\
    &= A_{t+1} \mathbb{E}_{q_\phi(x_{t+1}^K|\hat{y}_t)}\left[ (x_{t+1} - \mu_{x_{t+1}})(x_{t+1} - \mu_{x_{t+1}})^T\right] A^T_{t+1} \\
    &\approx A_{t+1} \frac{1}{N_s} \sum^{N_s}_i (\Tilde{x}^{(i)}_{t+1} - \mu_{\Tilde{x}_{t+1}})(\Tilde{x}^{(i)}_{t+1} - \mu_{\Tilde{x}_{t+1}})^T A^T_{t+1},
    \label{eq:sigma_calc}
\end{split}
\end{equation}
where $N_S$ denotes the number of drawn posterior samples $\Tilde{x}^{(i)}_{t+1}$.
Because the action space for $A_{t+1}$ scales with the binomial coefficient $N \choose M$, the computation is generally intractable and we instead explore only a subset of randomly selected actions $S_A$. We will refer to this policy as \textit{covariance sampling}: 
\begin{equation}
    A_{t+1}^* = \arg\max_{A_{t+1}\in S_A} \text{log det}(\Sigma_{y_{t+1}|A_{t+1}}).
    \label{eq:cov_sampling}
\end{equation}
Because this strategy is very computationally expensive, we propose an alternative sampling strategy that assumes independence across the expected observations, computing the trace of the covariance matrix instead. We will refer to this policy as \textit{trace sampling}:
\begin{equation}
    A_{t+1}^* = \arg\max \text{Tr}[\Sigma_{y_{t+1}|A_{t+1}}].
    \label{eq:trace_sampling}
\end{equation}

Since this method ignores the local correlation structure of closely-spaced ultrasound scan-lines (which originates from the limited physical resolution), we explicitly prohibit the system from choosing neighbouring lines. As an example; given a set of candidate actions [5,4,6,12,11,13] sorted by covariance traces that are of decreasing order, our approach samples lines 5 and 12 due to the neighbour-exclusion of lines 4 and 6. 
% For both sampling methods, the Markovian assumption \cite{Friedman2013AChange} holds for the posterior $z^K_t$ is inferred solely from observation $\hat{y}_{t-1}$. 
\section{Experiments \& Results}
\label{sec:exp}
%%%%%%%%%%%%%%%%%%%%%%%%%%%%%%%%%%%%%%%%%%%%%%%%%%%%%%%%%%%%%%%%%%%%%%%%%%%%%%%%%%%%%%%%%%%%%%%%%%%%%%%%%%%%%%%%%%%%%%%%%%%%%%%%%%%%%%%%%%%%%%%%%%%%%%%%%%%%%%%%%%%%%%%%%%
\subsection{Experimental Setup}
We evaluate our method using the EchoNet dataset \cite{Ouyang2020Video-basedFunction}, which contains 10,030 4-chamber cardiac ultrasound videos. Each video includes 50-250 grayscale frames with a resolution of $112\times 112$ pixels, captured at 50 Hz. To standardize the data, the pixel values are normalized to the range [0,1], and Gaussian noise $n\sim\mathcal{N}(0, 0.02)$ is added for improved generalization. The dataset is divided into 6,986 training videos, 500 validation videos for model selection, and 500 test videos for final evaluation. The remaining 2,044 videos are excluded due to artifacts or missing data.

We convert Cartesian images into polar coordinates, with depth $r = \sqrt{x_c^2 + y_c^2}$ and scan-line angle $\gamma = \arctan\left(\frac{y_c}{x_c}\right)$. To subsample full scan-lines, $A_t$ becomes highly structured and selects $M_\gamma< N_\gamma$ columns in the polar domain (i.e. $N=N_r N_\gamma$, $M=N_r M_\gamma$).

Our model architecture consists of a variational encoder, orthogonal Sylvester flow layers, and decoder. The encoder comprises 10 gated convolutional layers \cite{Dauphin2016LanguageNetworks} with stride 2, each using $c=64$ channels, reducing the input to a 512-dimensional latent variable $z_0=\mu + \sigma \cdot \epsilon$ with $\epsilon\sim\mathcal{N}(0, I)$, as per the reparameterization trick. The latent space is further refined using $K=8$ normalizing flow steps, each parameterized by $16$ orthogonal vectors ($N_p=16\times512)$, to obtain the final latent representation $z^K$. The decoder uses 8 blocks of gated transpose convolutions with stride 2, each using $c=128$ channels. These layers are followed by Batch Normalization \cite{Ioffe2015BatchShift} and GELU activations \cite{Hendrycks2016GaussianGELUs}. The final image is reconstructed through a head comprising 3 additional convolutional layers, each with $c=128$ channels.

We train the inference model with the loss function defined in \eqref{eq:VFE-Sylvester} and we set $\beta = 1 \times 10^{-4}$ for both the generative model and the inference model. We compensate for the polar coordinate transformation by using the density of the inverse transformation as a per-pixel weighing on the training loss. In an attempt to capture all modes for a given state of observation, the IWAE \cite{Burda2015ImportanceAutoencoders} algorithm is used, which tightens the ELBO.

We compare the two proposed sampling policies against three baseline methods: uniform random sampling, variable density random sampling, and equispaced sampling. In uniform random sampling, independent samples from a uniform distribution are used to select the $l$ columns for each frame. Variable density sampling uses a similar approach, but samples from a polynomial distribution centered on the middle of the image with a decay factor of 6 are used.  In the equispaced policy, the system deterministically uses evenly spaced lines and shifts the set of lines by one index for each subsequent video frame, maintaining uniform sampling density across all frames. For the trace and covariance-based sampling policies, we use $N_S=3$ posterior samples from our generative model and generate $S=10,000$ random sampling schemes to form the candidate set $S_A$ every $t$. Increasing $N_S$ and $S$ beyond these values resulted in increased computational costs with minimal performance improvement.

Although all subsampling methods share the same generative model, each has a distinct inference model. The training procedure is given in Algorithm~\ref{alg:training}. %It is important to note that once the system is deployed, only the active sampling component of the algorithm remains operational. 
The computational cost of the active methods is determined by the summation of the costs of the inference, sampling, image generation, and action selection steps.
%%%%%%%%%%%%%%%%%%%%%%%%%%%%%%%%%%%%%%%%%%%%%%%%%%%%%%%%%%%%%%%%%%%%%%%%%%%%%%%%%%%%%%%%%%%%%%%%%%%%%%%%%%%%%%%%%%%%%%%%%%%%%%%%%%%%%%%%%%%%%%%%%%%%%%%%%%%%%%%%%%%%%%%%%%

\begin{algorithm}
\caption{Training algorithm of the inference model for a single ultrasound video using active sampling.}\label{alg:training}
\begin{algorithmic}
\State \textbf{Require:} video $\boldsymbol{x} = [\hat{x}_1, ..., \hat{x}_T]$, empty subsampling matrix $A_t$, generative model parameters $\theta$, number of posterior samples $N_S$ and video length $T$.

\State $t \gets 0$
\While{$t \leq T$}
    \State $ \hat{y}_t \gets A_t\hat{x}_t + n_t$ \Comment{Following (\ref{eq:yaxn})}
    \State $ z^K \gets q_\phi(\hat{y}_t)$
    \State $ \Tilde{x}_t \gets p_\theta(z^K)$
    \State $ \Tilde{y}_t \gets A_t \Tilde{x_t} $
    \State \textbf{Back-propagate} $\mathcal{L}_{\Tilde{y}\rightarrow \hat{y}}$ + $\beta \mathcal{L}_{kld(z^K)}$
    \State \textbf{Optimizer step} $\phi$
    \State $\mu,\sigma, \lambda \gets q_\phi(\hat{y}_t)$ \Comment{Active sampling starts here}
    \State \textbf{Take $N_S$ samples $\boldsymbol{z^K}$ from $\mu,\sigma, \lambda$}
    \State $ \boldsymbol{\Tilde{x}_{t+1}} \gets p_\theta(\boldsymbol{z^K}) $
    \State $ A_{t+1}^* \gets \Sigma_{\Tilde{y}_{t+1}|A_{t+1},\hat{y}_t}  $ \Comment{Following (\ref{eq:cov_sampling}) or (\ref{eq:trace_sampling})}
    \State $t \gets t + 1$
\EndWhile
\end{algorithmic}
\end{algorithm}

%%%%%%%%%%%%%%%%%%%%%%%%%%%%%%%%%%%%%%%%%%%%%%%%%%%%%%%%%%%%%%%%%%%%%%%%%%%%%%%%%%%%%%%%%%%%%%%%%%%%%%%%%%%%%%%%%%%%%%%%%%%%%%%%%%%%%%%%%%%%%%%%%%%%%%%%%%%%%%%%%%%%%%%%%%
\subsection{Results}
\begin{table*}[ht]
    \centering
    \caption{Evaluation of the sampling strategies for different observation fractions.
    Performance is upper bounded by the generative model, under full observation L1-Loss=0.053, SSIM=0.523, PSNR=68.33.}
    \begin{tabular}{cccccccccccccccccc}
        \hline
          & & \multicolumn{3}{c}{l = 6 (5.4\% observation)} & \multicolumn{3}{c}{l = 9 (8.0\% observation)} & \multicolumn{3}{c}{l = 12 (10.7\% observation)} & \multicolumn{3}{c}{l = 15 (13.4\% observation)} \\ 
          & Active & L1-Loss & SSIM & PSNR & L1-Loss & SSIM & PSNR & L1-Loss & SSIM & PSNR & L1-Loss & SSIM & PSNR \\ \hline
        Variable density  &   No    & 0.086 & 0.401 & 65.53 & 0.078 & 0.428 & 66.09 & 0.073 & 0.450 & 66.44 & 0.070 & 0.456 & 66.75 \\ 
        Uniform random  &  No  & 0.085 & 0.396 & 65.64 & 0.076 & 0.435 & 66.31 & 0.069 & 0.457 & 66.78 & 0.065 & 0.476 & 67.21 \\ 
        
        Equispaced     &  No   & 0.073 & 0.447 & 66.54 & 0.064 & 0.477 & 67.28 & \textbf{0.060} & \textbf{0.494} & \textbf{67.73} & \textbf{0.058} & 0.502 & 67.89 \\

        Covariance (ours)  & Yes & 0.082 & 0.407 & 65.92 & 0.071 & 0.451 & 66.77 & 0.065 & 0.474 & 67.23 & 0.061 & 0.491 & 67.60 \\ 

        Trace (ours)& Yes & \textbf{0.070} & \textbf{0.455} & \textbf{66.69} & \textbf{0.062} & \textbf{0.495} & \textbf{67.51} & 0.061 & 0.489 & 67.60 & \textbf{0.058} & \textbf{0.500} & \textbf{67.93} \\ \hline
    \end{tabular}
    
    \label{table}
\end{table*}

\begin{figure*}[ht]
    \centering
    \includegraphics[width=\textwidth]{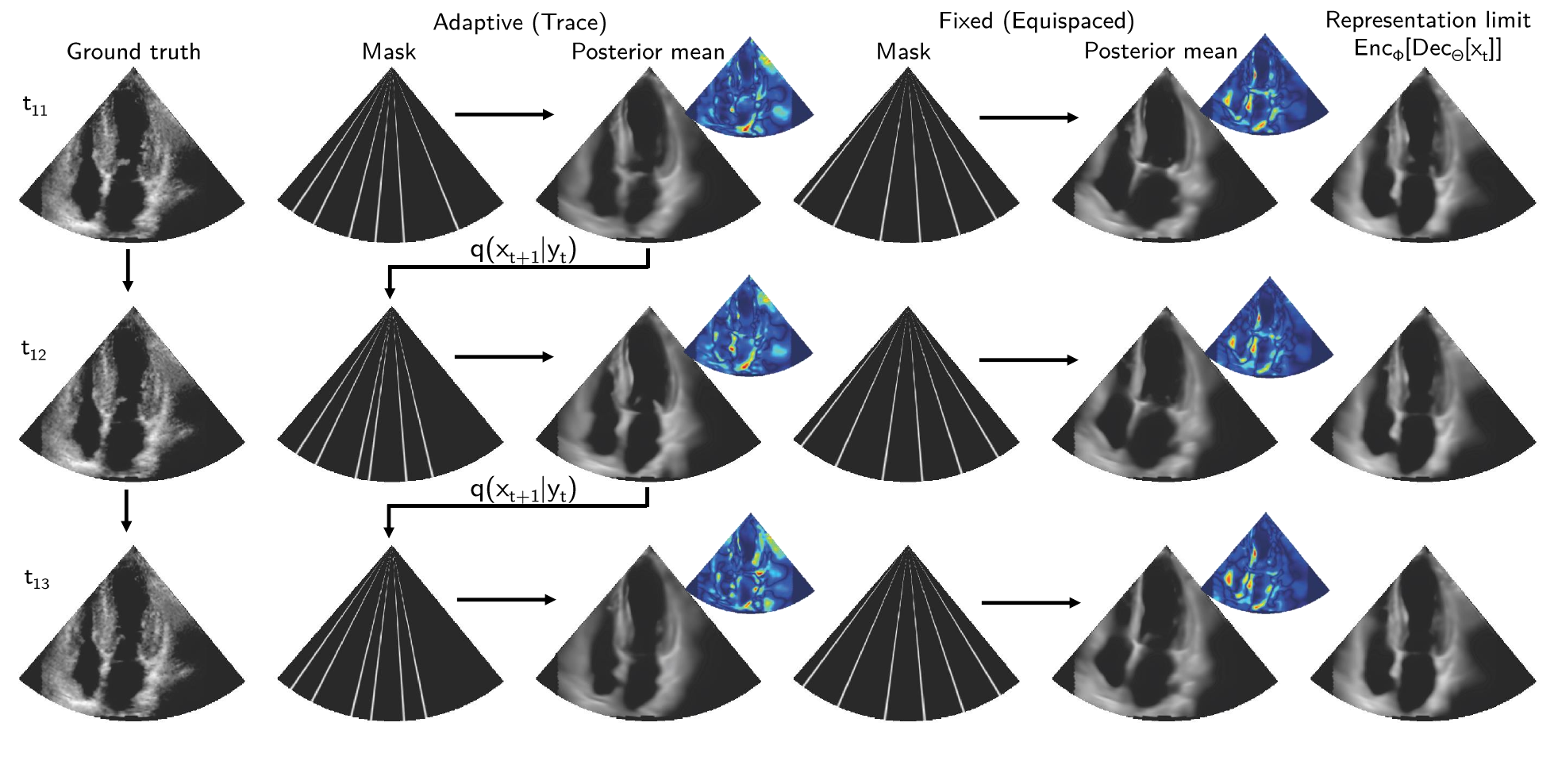}
    \caption{Reconstruction results for three consecutive frames $t_{11}, t_{12}$ and $t_{13}$ of an ultrasound video that has median performance for trace-sampling (L1-Loss = 0.070). The final column shows the representation limit that is given by the deep generative model (L1-Loss = 0.055). The smaller blue cones show the absolute difference between the posterior mean and the ground truth.}
    \label{fig:graphical_results}
\end{figure*}

Table \ref{table} presents the quantitative reconstruction results, evaluated using the L1-loss, Structural Similarity Index (SSIM), and Peak Signal-to-Noise Ratio (PSNR) across four different subsampling ratios. The proposed trace sampling method outperforms the other methods at subsampling rates of $l=6$, $l=9$, and $l=15$ scan-lines. However, for $l=12$ scan-lines, the equispaced sampling method performs slightly better. As the number of sampled lines increases, the performance gap between active and static sampling methods narrows, suggesting that active sampling is particularly advantageous when using more aggressive subsampling strategies. For equispaced and trace sampling, the upper bound for reconstruction is already approached with just $l=15$ lines (13.4\%).

A typical example (median performance) from the test set is visualized in Fig. \ref{fig:graphical_results}, illustrating the reconstruction results for three consecutive frames for the proposed trace sampling method with $l=6$ scan-lines.  For comparison, we also present the results using equispaced sampling and full sampling, which serves as the upper (representation) limit on performance. Since all methods share the same generative model, the differences in performance can be attributed to the sampling strategies only. As seen in the absolute-difference images, in this scenario the trace sampling favoured sampling the left side of the image for $t_{12}$ and $t_{13}$, leading to better reconstruction on the left at the expense of a slightly worse reconstruction on the right with respect to equispaced sampling. 
Interestingly, the trace-based sampling policy outperforms the covariance sampling method. % while outperforming the random methods, likely suffered from the Markov assumption. The masks that it suggested were optimized towards posterior updates, rather than posterior inference without memory. 

To assess the computational efficiency of our approach, we measured the time required for a complete acquisition step, including posterior estimation, on an NVIDIA GeForce RTX 2080 Ti (13.45 TFLOPS @ FP32), using the PyTorch 2.2 \cite{Paszke2019PyTorch:Library} backend. No additional optimizations were applied, such as JIT compilation, model pruning, or quantization. For trace sampling, a single acquisition step took 0.015 seconds, while covariance sampling required 0.112 seconds. This demonstrates that the proposed approach can operate at approximately 66 Hz, or potentially faster with further optimizations, making it suitable for real-time 2D ultrasound imaging applications.

% \begin{figure}[ht]
%     \centering
%     \subfigure[Equispaced (median L1-Loss)]{
%         \includegraphics[width=0.46\linewidth]{figures/equi-median-2.pdf}
%         \label{fig:suba}
%     }
%     \subfigure[Trace (median L1-Loss)]{
%         \includegraphics[width=0.46\linewidth]{figures/greedy-median-2.pdf}
%         \label{fig:subb}
%     }\\
%     \subfigure[Equispaced (10$^{th}$ \%ile L1-Loss)]{
%         \includegraphics[width=0.46\linewidth]{figures/equi-quartile-2.pdf}
%         \label{fig:subc}
%     }
%     \subfigure[Trace (10$^{th}$ \%ile L1-Loss)]{
%         \includegraphics[width=0.46\linewidth]{figures/greedy-quartile-2.pdf}
%         \label{fig:subd}
%     }
%     \caption{Visual results of median and worst $10^{th}$ percentile reconstruction for the equispaced and trace sampling methods. For every subplot; Top left: ground truth. Top right: reconstruction for full observation. Bottom left: $l=6$ lines of observation. Bottom right: corresponding reconstruction using the $l=6$ lines.}
%     \label{fig:graphical_results}
% \end{figure}

\section{Discussion and conclusion}
\label{sec:disc_and_concl}
In this paper, we proposed an active subsampling method for ultrasound scan-line selection that uses an information gain maximization policy in combination with a deep generative model and a neural posterior encoder. The results demonstrate that inference can be performed successfully at subsampling rates as low as 5.4\% and at frame rates of up to 66 Hz, making real-time active sampling feasible. Furthermore, we found that active sampling is especially beneficial under harsh subsampling regimes. 
This work opens up several avenues for future research. Firstly, because the sampling policy generates the observations on which the inference model is trained, and the inference model in turn affects the sampling policy, their optimization becomes intertwined. This may lead to collapse. %During training, the network is optimized towards reconstruction but does not have a metric that assesses the effectiveness of proposed sampling schemes. 
In addition, the influence of the $\beta$-parameter, which determines the trade-off between accurate reconstruction and diverse samples (both affecting the accuracy of the posterior proposition), should be studied. Alternatively two separate models could be trained; one to perform accurate maximum likelihood estimation for reconstruction, and one for posterior inference, driving sampling scheme generation. 

The proposed model also does not yet exploit long-term dependencies in the data, such as the cyclic nature of a beating heart. Future research to incorporate memory into the system, for example, through the use of self-attention \cite{Vaswani2017AttentionNeed} or LSTM \cite{Beck2024XLSTM:Memory}, could further improve reconstruction results and/or lead to even more aggressive subsampling schemes. Additionally, the use of a more powerful deep generative model, such as the VD-VAE \cite{Child2021VeryImages}, would lead to a more accurate posterior approximation, improving the active subsampling schemes even further.
Lastly, the results on 2D ultrasound show promise for application in 3D ultrasound, where the trade-off between volume rate and image resolution is far more challenging to manage.

\bibliographystyle{IEEEtran}
\bibliography{references}

\end{document}